\documentclass[twocolumn,aps]{revtex4}

\usepackage{amsmath,amssymb}
\usepackage{graphicx}
\usepackage{dcolumn}
\usepackage{bm}

\begin{document}

\preprint{APS/123-QED}

\title{On the energetic origin of self-limiting trenches formed 
 around Ge/Si quantum dots}

\author{D.~T.~Tambe}
\author{V.~B.~Shenoy}
\email{shenoyv@engin.brown.edu}
\affiliation {Division of Engineering, Brown University, Providence, RI 02912}

\date{\today}

\begin{abstract}
At high growth temperatures, the misfit strain at the boundary 
of Ge quantum dots on Si(001) is relieved by formation of
trenches around the base of the islands. The depth of the trenches has been observed to saturate at a level that depends on the base-width of the islands. Using
finite element simulations, we show that the self-limiting nature of 
trench depth is due to a competition between the elastic relaxation energy gained by the formation of the trench and the surface energy cost for creating the trench.
Our simulations predict a linear increase of the trench depth with the island radius, in quantitative agreement with the experimental observations of Drucker and coworkers \cite{jeff1}.

\end{abstract}

\maketitle

Heteroepitaxial growth of strained semiconductor thin films has been a subject of interest in the recent years, as it provides a versatile route for fabricating nanostructures that have potential applications as electronic, optoelectronic or memory devices. Lattice mismatch is one of the key elements that has been exploited  to grow structures such as quantum dots and nanowires. Strain-driven self assembly in Ge/Si(001) has received particular attention, as 3D islands formed in this system can be integrated as devices that are compatible with well-developed Si technology. It is well known that mismatch strain at low Ge coverages is relaxed by the formation of pyramid-shaped islands with \{105\} facets. As these islands grow in size, they transform into what are called domes with steeper sidewall angles. Even larger islands show strain relaxation through the injection of misfit dislocations. In high temperature growth (T $>$ 600$^0$C), strain relaxation is also achieved by trenches that form around the bases of both coherent and dislocated islands \cite{jeff1,jeff11,jeff2,jeff3,denker1,denker2,kamins,floro}. 

Experimental work of Chapparo and coworkers \cite{jeff1} has shown that the trenches at the base of dome-shaped islands are self-limiting, in the sense that their depths saturate at a level, linearly proportional to the base-width of the island. While it has been suggested that this behavior is due to kinetic limitations \cite{jeff1}, quantitative estimates for the self-limiting depth and a reason for its linear dependence of the island size have not been provided. In this article, we consider the energetics and strain relaxation of islands surrounded by trenches using finite element simulations and show that the self-limiting nature of trenches can be explained on the basis of a competition between the elastic energy gained by the formation of the trenches and the surface energy cost involved in creating the trenches. Our calculations that employ trenches whose dimensions and shapes are comparable to the ones observed in experiments, predict a linear dependence of the trench depth on the island base-width, in good quantitative agreement with experimental data.


\begin{figure}
\begin{center}
{\resizebox{!}{5.cm}{\includegraphics{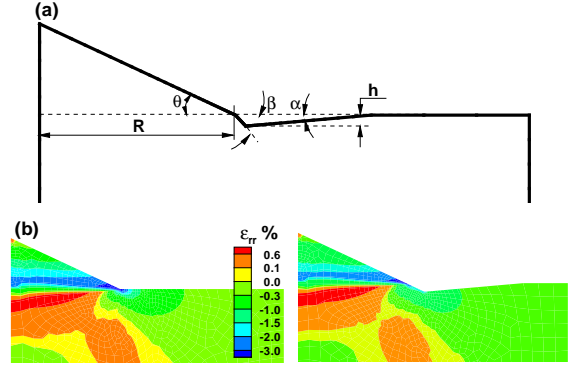}}}
\caption{\label{fig:model} (a) Schematic of the axisymmetric model of a cone-shaped Ge/Si island of radius $R$ and 
cone angle $\theta$. The depth of trench around the base of the island is denoted by $h$ and the trench angles $\beta$ and $\alpha$ are shown in the figure.
(b) Contour plots of distribution of 
radial strain $\varepsilon_{rr}$ for an island with (right) and without trench (left).
Here $h =$ 10nm, $R=125$, $\theta$=$\beta=$25$^0$ and $\alpha$=5$^0$.}
\end{center}
\vspace{-0.5cm}
\end{figure}
We first consider strain relaxation due to trench formation around a cone shaped island shown in Fig.~\ref{fig:model}(a). The cone radius and angle are denoted by $R$ and  $\theta$, respectively, while a trench of depth $h$ is modeled using two inclined surfaces at angles $\beta$ and $\alpha$ shown in Fig.~\ref{fig:model}(a). In the case of trenches observed in experiments \cite{jeff1,jeff11,denker1}, $\beta$ is comparable to or bigger than the island sidewall angle $\theta$, while the angle $\alpha$ is much smaller, usually a few degrees.
A displacement based finite element method with linear shape functions is used to obtain
the elastic fields in the island and the substrate.
Strain in the structure can be expressed as
\begin{equation}
\varepsilon_{ij} = \varepsilon^0_{ij} + u_{ij}
\nonumber
\end{equation}
where, $\varepsilon^0_{ij}$ is the mismatch strain (present only in the island) and
$u_{ij}$ is the {\em additional} strain due to elastic relaxation. The non-zero components of the relaxation stress tensor (in cylindrical coordinates) are given by
\begin{equation}
\left(
\begin{array}{c}
\sigma_{rr}\\ \sigma_{\theta \theta}\\ \sigma_{zz}\\ \sigma_{rz}\\
\end{array}
\right) =
\left(
\begin{array}{cccc}
C_{11} & C_{12} & C_{12} & 0\\
C_{12} & C_{11} & C_{12} & 0\\
C_{12} & C_{12} & C_{11} & 0\\
0 & 0 & 0 & (C_{11}-C_{12})/2\\
\end{array}
\right)
\left(
\begin{array}{c}
u_{rr}\\ u_{\theta \theta}\\ u_{zz}\\ u_{rz}\\
\end{array}
\right),
\nonumber
\end{equation}
where the elastic constants of the Si substrate and the Ge island are taken to be C$_{11}$ = 165.7~GPa, 
C$_{12}$ = 63.9~GPa and C$_{11}$ = 128.5~GPa, C$_{12}$ = 48.3~GPa, respectively. If the strain fields are obtained from finite element calculations, the free energy of the system, which is the sum of the energy gained through elastic relaxation and the increase in surface energy due to formation of the trench, can be written as
\begin{equation}
F= - \frac{1}{2} \int_{V_S} \sigma_{ij} u_{ij} dV + \gamma \Delta A.
\label{eq:freeenergy}
\end{equation}
Here, the first integral is evaluated over the volume $V_S$ of the system, which includes the island and the substrate with the trench, $\Delta A$ denotes the increase in surface area due to formation of the trench and $\gamma$ is the surface energy of the substrate, assumed to be independent of strain and orientation \cite{surfaceenergy}.

The radial strain distribution in the island with and without the trench for $\beta = \theta = 25^0$, $\alpha = 5^0$ and an equibiaxial mismatch strain of $-4 \%$ is shown in Fig~1(b).
In both the cases, the radial strain $\varepsilon_{rr}$
changes sign across the island substrate interface
with the substrate side under tension. The apex of the island 
is slightly over-relaxed \cite{spencer}.
The key difference between the case with the trench compared to the case without the trench relates to the strain distribution around the base of the island. The presence 
of trench leads to an effective relaxation of the large near-surface compressive 
strain at point where the island makes contact with the substrate. At the same time, the trench also allows relaxation of tensile strain near the substrate surface. 
The azimuthal strain $\varepsilon_{zz}$
distribution has a maximum compressive value below the island
boundary, so that the substrate near the boundary of the island tends to
relax upwards and the material in the island tries to relax radially
outwards.

\begin{figure}
\begin{center}
\includegraphics[width=8.2cm ]{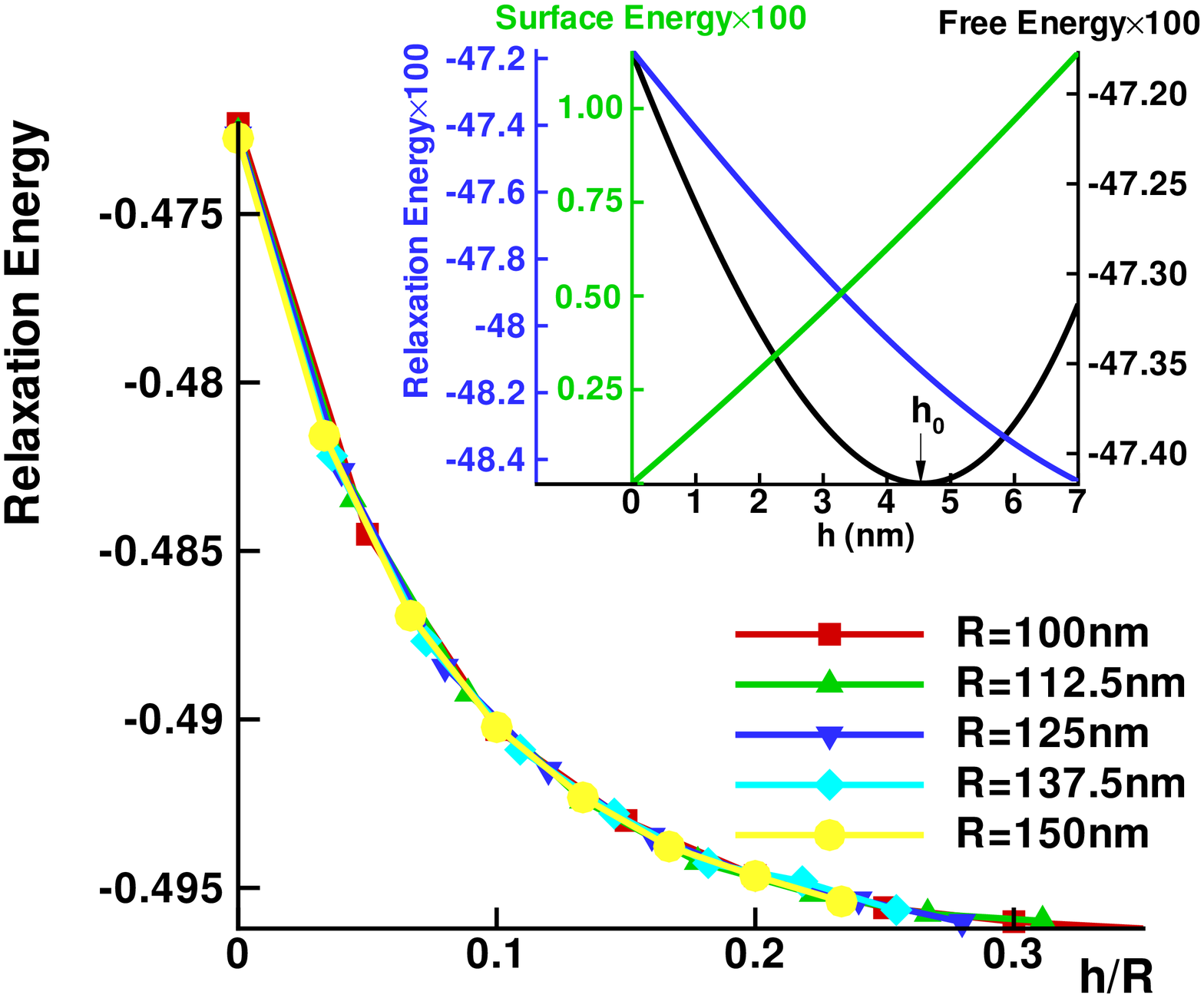}
\vspace{-0.5cm}
\caption{\label{fig:energy} Elastic relaxation energy of an island with a trench normalised by the total strain energy of the island prior to relaxation $M\epsilon_0^2V$, plotted versus the ratio of trench depth $h$ to island radius $R$.
Note that the relaxation energy attains saturation at $h/R \sim $ 0.3. The surface orientations (refer to Fig.~1) in this calculations are $\alpha$=5$^0$ and $\beta$=$\theta$=25$^0$.
The inset shows the competition between the elastic relaxation energy (blue) and 
the change in surface energy (green, normalised by $M\epsilon_0^2V$) resulting in a minimum in the total free energy (black) at an optimum trench depth ($h_0$) for
an island with $R =$ 125 nm and $\gamma$=1.5~J/m$^2$.}
\end{center}
\vspace{-0.7cm}
\end{figure}

The elastic relaxation energy as a function of trench depth $h$ is given in Fig.~\ref{fig:energy} for islands of different radii, but fixed surface orientations $\theta$, $\beta$ and $\alpha$. Here, we have plotted the first term in Eq.~(\ref{eq:freeenergy}) normalized by the total strain energy of the island prior to relaxation, $M \epsilon_0^2 V$, where $M$ is the biaxial modulus \cite{biaxial}, $\epsilon_0 = -4\%$ is the mismatch strain and $V = \pi R^3 \tan \theta /3$ is the volume of the island. As the trench depth is increased, the elastic energy initially decreases at a rapid rate, but eventually saturates at $h/R \sim 0.3$. Since the strain fields decay rapidly from the island substrate interface, the energy gained by extending the trenches to relatively unstrained regions is small, leading to the observed saturation in strain relaxation. We also find that normalized energies for islands of different radii fall on a single curve if the trench depth is scaled with the radius of the island, which shows that deeper trenches are required to achieve saturation of elastic energy for bigger islands. While elastic energy is relaxed by increasing trench depth, deeper trenches also lead to an increase in the total surface energy of the system. The inset in Fig.~2 shows that the net change in surface energy is comparable to the energy gained via strain relaxation for an island with $R = 125$ nm. Furthermore, a competition between these two effects leads to a self-limiting trench size, $h_0$,  which corresponds to the minimum in the free energy $F$ (refer to Eq.~2) as shown in the inset in Fig.~2. We point out that while the gain in elastic energy by formation of trenches is only 2\% of the total strain energy of the structure without trench, this corresponds to 3.55$\times$10$^5$$k_BT$ at 600$^0$C for the island in Fig.~2; the trenches should then be very stable against thermal fluctuations. The net gain in free energy and the self-limiting trench depth depends on the surface orientations of the island and the trench and the size of the island. We consider these effects in the following paragraph.

\begin{table}
\caption{\label{tab:table1} Effect of variations in the surface orientations $\beta, \alpha$ and $\theta$ on  
the optimum trench depth ($h_0$ in nm) and the normalized relaxation energy 
($\Delta F \times 10^{-3}$, where $\Delta F = F(h=0)-F(h=h_0)$) with respect to the ungrooved structure.
When the variations of a particular parameter is considered, the default values of other parameters are, 
$\theta=\beta$=25$^\circ$, $\alpha$=5$^\circ$ and $\gamma$=1.5~J/m$^2$.}
\begin{tabular}{||ccc||ccc||ccc||}
\hline
$\beta$ & $\Delta F$ & $h_0$ & $\alpha$ & $\Delta F$ & $h_0$ & $\theta$=$\beta$ & $\Delta F$ & $h_0$ \\ \hline
25$^\circ$ & 2.21 & 4.56 & 5$^\circ$ & 2.21 & 4.56 & 15$^\circ$ & - & -\\ \hline
33$^\circ$ & 2.48 & 4.92 & 10$^\circ$ & 0.62 & 3.0 & 20$^\circ$ & 0.23 & 2.50 \\ \hline
47$^\circ$ & 2.83 & 5.51 & 15$^\circ$ & 0.19 & 1.42 & 25$^\circ$ & 2.21 & 4.56 \\ \hline 
\end{tabular}
\vspace{-0.5 cm}
\end{table}

The data in the first column of Table~\ref{tab:table1} shows that the optimum trench depth $h_0$ increases by about 1nm when the trench angle $\beta$ is increased from 25$^0$ to 47$^0$ at fixed $\theta$ and $\alpha$ for $R =$ 125nm. We can understand this trend by noting that larger $\beta$ leads to more effective relaxation of the compressive strain at the point where the island meets the substrate and the tensile strain in the substrate. While the free energy of the system can be lowered for larger $\beta$, the opposite is true for the trench angle $\alpha$ (refer to the second column of Table~\ref{tab:table1}). Lower values of  $\alpha$ allow better relaxation of the tensile strain in the substrate and involve a smaller change in surface energy. This result is in agreement with experimental observations \cite{jeff1,jeff2,jeff11,denker1,denker2} that show steep (large $\beta$) and shallow (small $\alpha$) surface orientations of the trench close to and away from the island, respectively. If the sidewall angle of the island is lowered by choosing $\beta = \theta$ and by holding $\alpha$ constant, optimum trench depth $h_0$ decreases in a monotonous manner (refer to the third column in Table.~\ref{tab:table1}) until it becomes unfavorable to form trenches for $\theta \le 15^0$. This result would indicate that the 
trenches would not be observed at around the hut-shaped \{105\}-faceted islands whose base-widths are $\approx$ 200 nm,  in agreement with experimental observations. Next, we consider the evolution of the trench depth with the radius of the island, with the aim of comparing the predictions of our calculations with the data reported in Ref.~1.

\begin{figure}
\begin{center}
\includegraphics[width=8.2cm ]{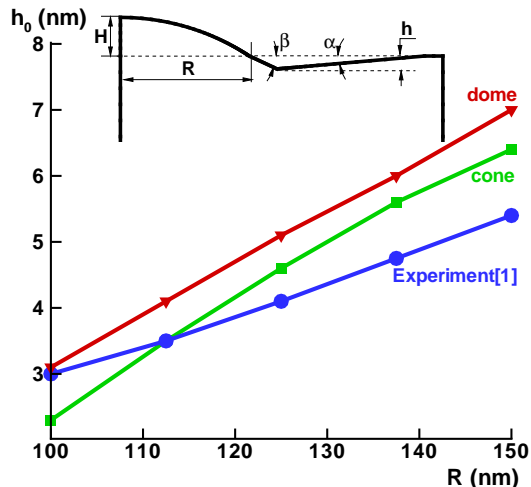}
\vspace{-0.5cm}
\caption{\label{fig:saturation} Self-limiting trench depth versus island diameter for a conical
Ge/Si island with $\theta$=$\beta$=25$^0$, $\alpha$=5$^0$ and for the 
dome shaped island shown in the inset with $H/R =$ 0.3 and $\beta$=25$^0$, $\alpha$=5$^0$. The surface energy is $\gamma$=1.5 J/m$^2$ in both the cases.
Saturation depth observed in the experiments \cite{jeff1} is shown by the filled circles.}
\end{center}
\vspace{-0.8cm}
\end{figure}

Since the AFM and TEM images of the cross-sections of the domes \cite{jeff1,jeff11,jeff2} indicate that the steeper face of the trench is at the same  orientation as the sidewall of the dome cluster, we take $\beta = \theta =$ 25$^0$, which is very close to the angles made by the \{113\} and \{102\} facets with the substrate. The dependence of the optimum trench depth as a function of the island radius is given in Fig.~\ref{fig:saturation} for the conical island in Fig.~1 and for a dome-shaped island shown in the inset in Fig.~3. The axially symmetric sidewall of the latter shape is an arc of a circle with an aspect ratio (ratio of height $H$ to radius $R$) of 0.3. In both cases, we find nearly linear increase of the optimum trench depth with the island radius, which is in good quantitative agreement with experimental data from Ref.~1.  It should be pointed out that our calculations predict somewhat larger slopes for the curves, which could be attributed to a couple of modeling assumptions. In our calculations, we have taken that the composition of the islands is uniform (100\% Ge), while there is experimental evidence \cite{jeff2} that Si diffuses from the substrate into the island at temperatures considered in Ref.~1. In this case, since the elastic energy gain would be smaller than our estimates, the trench depths would also be somewhat smaller, bringing our results closer to the experimental data. Closer agreement could perhaps be achieved if  strain and orientation dependence of surface energy is included in the analysis.

In conclusion, we have shown that the self-limiting behavior of the trenches around the bases of SiGe quantum dots can be explained on the basis of a competition between elastic relaxation and the surface energy cost of creating the trench surface. The optimum trench depth has an approximate linear dependence with the island radius, due to the fact that more elastic energy can be relaxed by trench formation in the case of larger islands. Our results indicate that the saturation of the trench depths can be attributed to energetics rather than  kinetic limitations on mass transport. 

The research support of the National Science Foundation
through grants CMS-0093714 and CMS-0210095 and the Brown University
MRSEC program through grant DMR-0079964 is gratefully acknowledged.


\begin {thebibliography}{30}
\bibitem{surfaceenergy} Since data on the strain and orientation dependence of surface energies is not available from experiments or accurate quantum mechanical calculations, we use $\gamma =$ 1.5 J/m$^2$, which is typical for semiconductor surfaces.
\bibitem{biaxial} The biaxial modulus can be written in terms of the elastic constants as $M = C_{11}+C_{12}-2C_{12}^2/C_{11}$.
\end{thebibliography}
\end{document}